\documentclass[10pt,sigconf]{acmart}
\usepackage{graphicx}
\usepackage{balance}  
\usepackage{pgfplots}
\usepackage{booktabs}
\usepackage{url}
\usepackage[ruled,vlined,linesnumbered]{algorithm2e}
\usepackage{algorithmic}

\usepackage{booktabs}
\usepackage{tabularx}
\usepackage{rotating}
\usepackage{paralist}
\usepackage{subfigure}
\usepackage{stmaryrd}
\usepackage{datetime}
\usepackage{balance} 
\usetikzlibrary{datavisualization} 
\usetikzlibrary{datavisualization.formats.functions} 
\usepackage{graphicx,multicol}
\usepackage{pgfplots}
\usepgfplotslibrary{groupplots}
\pgfplotsset{compat=1.12}
\usetikzlibrary{patterns}
\usepackage{tikz}
\usetikzlibrary{calc,shapes.geometric,arrows,fit,matrix,positioning,pgfplots.groupplots,trees,patterns}
\AtBeginDocument{%
  \providecommand\BibTeX{{%
    \normalfont B\kern-0.5em{\scshape i\kern-0.25em b}\kern-0.8em\TeX}}}

\hypersetup{draft}

\setcopyright{none}
\hypersetup{draft}

\begin{document}

\settopmatter{printacmref=false} 
\renewcommand\footnotetextcopyrightpermission[1]{} 
\pagestyle{plain} 

\title{AIR -- A Light-Weight Yet High-Performance Dataflow Engine based on Asynchronous Iterative Routing}

\author{Vinu E. Venugopal}
\author{Martin Theobald}
\author{Samira Chaychi}
\author{Amal Tawakuli}
\email{{vinu.venugopal,martin.theobald,samira.chaychi,amal.tawakuli}@uni.lu}
\affiliation{
  \institution{Faculty of Science, Technology and Communication}
  \city{University of Luxembourg, Belval}
  \country{Luxembourg}
  \postcode{4364}
}


\renewcommand{\shortauthors}{Venugopal et al.}

 \newcommand{\squishlist}{
 \begin{list}{$\bullet$}
  { \setlength{\itemsep}{0pt}
     \setlength{\parsep}{1pt}
     \setlength{\topsep}{1pt}
     \setlength{\partopsep}{0pt}
     \setlength{\leftmargin}{1em}
     \setlength{\labelwidth}{1em}
     \setlength{\labelsep}{0.5em} } }
 \newcommand{\squishend}{\end{list}}
 
\begin{abstract}
Distributed Stream Processing Systems (DSPSs) are among the currently most emerging topics in data management, with applications ranging from real-time event monitoring to processing complex dataflow programs and big data analytics. The major market players in this domain are clearly represented by Apache Spark and Flink, which provide a variety of frontend APIs for SQL, statistical inference, machine learning, stream processing, and many others. Yet rather few details are reported on the integration of these engines into the underlying High-Performance Computing (HPC) infrastructure and the communication protocols they use. Spark and Flink, for example, are implemented in Java and still rely on a dedicated master node for managing their control flow among the worker nodes in a compute cluster. 

In this paper, we describe the architecture of our AIR engine, which is designed from scratch in C++ using the Message Passing Interface (MPI), pthreads for multithreading, and is directly deployed on top of a common HPC workload manager such as SLURM. AIR implements a light-weight, dynamic sharding protocol (referred to as ``Asynchronous Iterative Routing''), which facilitates a direct and asynchronous communication among all client nodes and thereby completely avoids the overhead induced by the control flow with a master node that may otherwise form a performance bottleneck.
Our experiments over a variety of benchmark settings confirm that AIR outperforms Spark and Flink in terms of latency and throughput by a factor of up to 15; moreover, we demonstrate that AIR scales out much better than existing DSPSs to clusters consisting of up to 8 nodes and 224 cores.
\end{abstract}

\keywords{Distributed stream processing systems; Dataflow programming; Stateful windowed operators; Asynchronous routing}
\maketitle

\section{Introduction}

With the recent advent of Industry 4.0 and Internet-of-Things (IoT) applications on a broad basis, the scalable processing and real-time analysis of streaming data has reached an unprecedented attention both in academia and industry. Catching up with the vastly growing rate at which data streams are produced already today poses a major---if not intangible---challenge to classical database architectures and is therefore increasingly replaced by distributed analytical platforms based on Apache Hadoop \cite{hadoop}, Spark \cite{SStreaming}, Storm \cite{storm} and Kafka \cite{kafka}. This new generation of tools is designed to process data streams in a flexible, scalable, fast and resilient manner, but still is largely limited to splitting an incoming data stream into batches and to then synchronously execute their analytical workflows over these data batches. To overcome the limitations of this iterative form of {\em bulk-synchronous processing} (BSP), {\em asynchronous stream-processing} (ASP) engines such as Apache Flink~\cite{FlinkCarboneKEMHT15}, Samza \cite{samaza} and Naiad~\cite{mcsherry2013differential,Murray:2013} have recently emerged and are meanwhile increasingly being fostered both by open-source communities, such as the {\em Apache Software Foundation} (ASF), and by a new line of companies like {\em Databricks} and {\em dataArtisans}.

What all of the aforementioned platforms (both BSP and ASP) still have in common, though, is their rather static {\em master-client architecture}, which still dates back to the original design of Hadoop. 
Given the inherent deviances of distributed computations (due to communication and network delays, scheduling algorithms, time spent on processing, serialization/deserialization, etc.), the performance of platforms built on such a master-client architecture is still bound by hidden synchronization barriers and the constant need of state exchange (and hence communication overhead) between the master and the worker nodes~\cite{Verma:2013}.
To tackle this prevalent bottleneck of existing master-client architectures, we present a novel ASP architecture that is {\em completely client-client based}. Our engine, called AIR, is based on a novel communication protocol among the worker nodes, which we refer to ``Asynchronous Iterative Routing'', to process one or more incoming data streams in a completely asynchronous manner. AIR has been designed from scratch in C++ and is purely based on the Message Passing Interface (MPI), which provides a low-level and highly efficient layer of communication among the workers. 
It, therefore, does not inherit any legacy design of an existing platform and can freely be adapted to our needs. It is neither restricted to the rather static, receiver-initiated message passing strategy employed by Hadoop and Flink~\cite{tcp} nor to the actor-based Akka API used by Spark~\cite{akka}, but allows for a highly multithreaded communication among the worker nodes directly via channels. AIR is available as open-source release at  our GitLab repository\footnote{\url{https://gitlab.uni.lu/mtheobald/AIR}}.

\begin{figure}[htb]
\centerline{\includegraphics[width=0.48\textwidth]{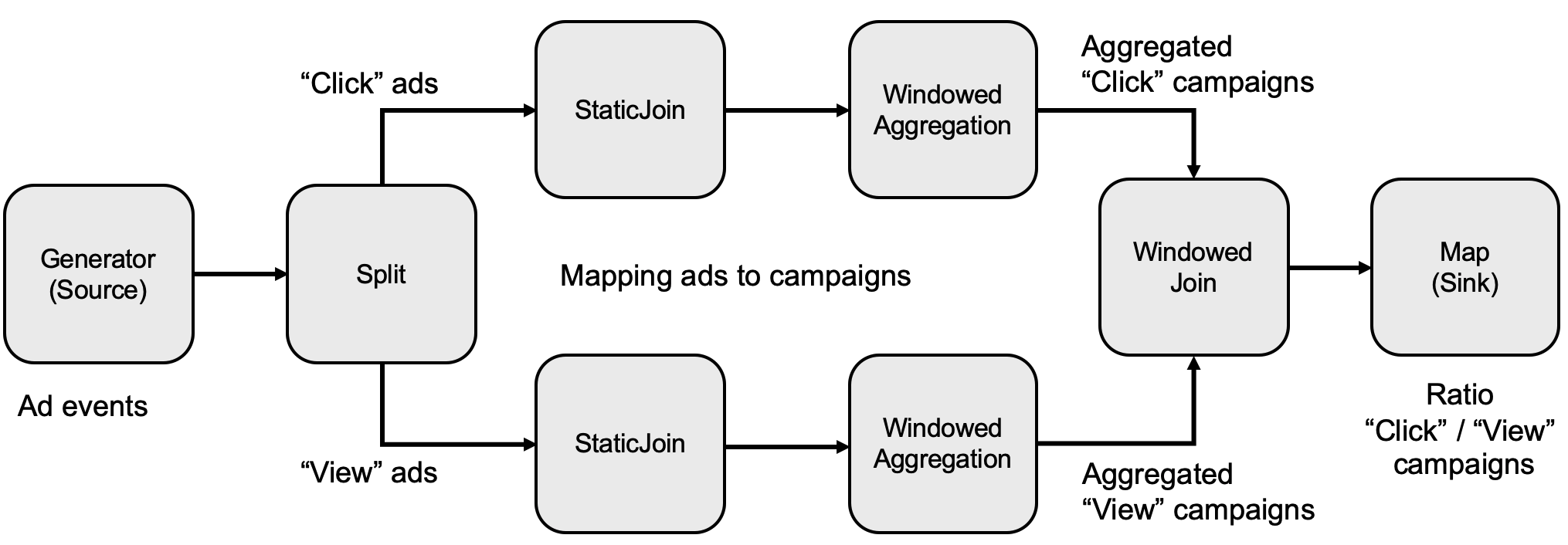}}
\caption{Extended Yahoo Streaming Benchmark (YSB)}
\label{fig:extendedYSB}
\end{figure}
\textit{Benchmark Setting: Figure~\ref{fig:extendedYSB} depicts an extended version of the Yahoo Streaming Benchmark (YSB)~\cite{DSPBM18}, in which we compute the ratio among ``Click'' and ``View'' events of ads belonging to various campaigns over a given sliding window (e.g., 10 seconds). As opposed to the original YSB setting, which uses only a single sliding-window aggregation, we first split the generated events into two ``Click'' and ``View'' streams, which are then statically joined with their respective campaign ids and processed by two sliding-window aggregations separately.  A third sliding-window join then computes the ratio among the two aggregated event streams over each window. AIR outperforms Spark by a factor of 4.3 and Flink by a factor of 3.4 at a maximum sustainable throughput of 269 million events per second (corresponding to 36.58 GB/sec.) on a cluster of 8 compute nodes and an overall amount of 224 cores in this setting.}

\subsection{AIR Overview \& Contributions}
Our contributions are a new, highly multithreaded, distributed and asynchronous dataflow processing model, along with an implementation of the AIR framework, which allows for a direct HPC deployment based on any underlying MPI library (such as MPICH or OpenMPI) and workload manager (such as SLURM or OpenStack).
\squishlist

\item \textbf{Master-less architecture}: the overhead of maintaining a dedicated master node, which periodically polls the worker nodes for their current state and workload, is completely avoided by a dynamic sharding protocol called ``Asynchro\-nous Iterative Routing''.

\item \textbf{Globally asynchronous transformation operators}:\\ stateless operators, such as {\tt Map}, {\tt Split} and {\tt Filter}, process their tasks in a completely asynchronous manner both (1) across the communication channels per rank as well as (2) across the ranks within a cluster.

\item \textbf{Locally asynchronous sliding-window operators}:\\ stateful operators, such as {\tt Reduce}, {\tt Join}, and {\tt Aggregate}, use a combination of (1) asynchronous local preprocessing and (2) synchronous global processing, which thus globally need to synchronize only per sliding window (to guarantee correctness) but overall remain asynchronous across different windows.

\item \textbf{Multithreaded channel processing}: communication\\ channels among ranks are implemented on top of the MPI APIs in a highly multithreaded manner, thus contributing to an increased core utilization per dataflow operator.

\item \textbf{Pipelining:} for stateless operators, such as {\tt Map}, {\tt Split} and {\tt Filter}, MPI communication may optionally be reduced by a direct form of pipelining messages between a sending thread's outgoing message queue and a receiving thread's incoming message queue.


\squishend

\section{Background \& Preliminaries}

Before we introduce the detailed architecture of AIR, we briefly review the main design differences among the major DSPS platforms, which are most notably represented by Apache Flink and Spark as open-source engines. While Flink (as the successor of Stratosphere \cite{stratosphere}) has been designed from scratch for stream processing, the roots of Spark clearly are in batch processing. Only as of Spark 2.3, a notion of ``continuous streams'' \cite{ApacheSparkSS} has been introduced to Spark's principal architecture. We therefore still list Spark as a bulk-synchronous architecture, although the exact boundaries among the two paradigms are increasingly vanishing. In addition to the DSPSs listed in the following two sections, there exist a plethora of more systems~\cite{Venkataraman:2017,Lin:2016,Arasu04stream,Akidau:2013,Kulkarni:2015,Buddhika2016NEPTUNERT,AbadiABCCHLMRRTXZ05,Thies:2002:SLS,Gedik:2008}, some of which are specifically designed to exploit HPC-style multi-core architectures~\cite{MisaleDTMA18,Streambox,fflow}.

\subsection{Bulk-Synchronous Processing (BSP)}

BSP engines 
divide a potentially unbounded stream of input data into manageable batch sizes before performing their logical operations on these batches. BSP systems therefore need to synchronize their computations between each two subsequent batches. Also, sliding-window operators, such as aggregations or joins, are aligned with the batch sizes, such that the consideration of time-stamps and out-of-order events within these operators are not a principal concern. 

\smallskip
\noindent\textbf{Apache Storm} is a real-time DSPS written in Java and Clojure~\cite{storm}, which follows a classical master-client architecture. Once the topology of the application is defined in the form of a DAG of operators, the master specifies the computation pipeline and schedules the tasks to the workers. In addition, the master has to coordinate each worker process by considering this static topology specification. The worker nodes monitor their processes and periodically send their current state to the master by using a so-called ``heartbeat'' protocol. 

\smallskip
\noindent\textbf{Apache Spark} is a major platform for distributed computations (including stream processing) written in Java and Scala~\cite{SStreaming}. Distributed  computations are based on Resilient Distributed Datasets (RDDs), which essentially are distributed in-memory data structures. Spark Streaming is a major extension of the core Spark APIs, in which an input stream is treated as a sequence of micro-batches, then called a {\em DStream} (i.e., a sequence of individual RDDs) \cite{ApacheSparkSS}.  Also Spark follows a master-client architecture, where a \emph{driver} program is responsible for creating the Spark Context and for scheduling tasks across the worker nodes (then called ``task managers''). The Spark Context connects to the underlying cluster manager which is responsible for resource allocation between applications. Once connected, Spark executes tasks within the task managers, which perform processing and data storage (largely equivalent to Storm workers), and communicate their results back to the Spark Context.  

\smallskip
\noindent\textbf{Thrill} is a C++ prototype of a general-purpose BSP architecture developed at KIT (Karlsruhe)~\cite{DBLP:conf/bigdataconf/BingmannAJLNNSS16}. Similarly to Spark, Thrill's data model uses Distributed Immutable Arrays (DIAs), in which operations are only permitted to the array as a whole. DIAs act as abstract entities flowing between the operations, thus allowing for optimizations such as pipelining or chaining. Thrill provides an SPMD (Single Program, Multiple Data) execution model, similar to MPI, where the same program is executed on different machines in parallel.

\subsection{Asynchronous Stream Processing (ASP)}

Recently, ASP engines have been designed in the spirit of a so-called {\em kappa-architecture} \cite{kappa14}, in which the granularity of an incoming data stream is largely independent of the granularity of the analytical tasks performed over the stream. Although, also here, an incoming data stream is usually discretized into conveniently sized messages with respect to the underlying network protocol (such as TCP-IP), the aggregations performed over a sliding window of incoming messages are largely decoupled from the actual message sizes. With an ASP architecture, the consideration of individual event-times and the handling of out-of-order events plays a key role for the correct execution of sliding-window operators, such as aggregations or joins. 

\smallskip
\noindent\textbf{Apache Flink} is a hybrid processing platform written in Java and Scala~\cite{FlinkCarboneKEMHT15}. The core of Flink is a stream processing engine, thus allowing for batch processing as a special case. Like most DSPSs, Flink follows a master-client architecture. Upon receiving a job, the job manager (i.e., the master) generates the corresponding tasks and assigns them to the workers. In addition, the job manager maintains the state of all executions and the status of all worker nodes. Task Managers (i.e., the workers) perform tasks assigned by the job manager and exchange information with other workers when needed. Each task manager provides a number of processing \emph{slots} to the cluster which are then used to perform tasks in parallel. The abstraction for streams in Flink is called \emph{DataStream}, which is a sequence of partially ordered records.

\smallskip
\noindent\textbf{Apache Samza} \cite{samaza} a distributed stream processing system that supports stateful operators and adopts a unified design for both real-time stream as well as batch processing along the given dataflow structure. The overall framework is tightly coupled with the Apache Kafka messaging system. While Kafka can be used by many stream processing systems, Samza is designed specifically to take advantage of Kafka's infrastructure regarding fault tolerance, buffering, and persistency. 

\smallskip
\noindent\textbf{Naiad}~\cite{mcsherry2013differential,Murray:2013} is a distributed high-performance dataflow framework developed at Microsoft Research. The core of Naiad is implemented in C\#. It follows a master-client architecture where workers exchange messages locally using shared memory, and remotely using a TCP connection. It follows a new computational model called ``timely dataflow'', which extends traditional incremental computations to allow arbitrarily nested iterations. Moreover, it enriches dataflow computations with timestamps that represent logical points in the computations which provide the basis for an efficient, lightweight coordination mechanism.

\smallskip
\noindent The \textbf{Dataflow} \cite{43864} and \textbf{Millwheel} \cite{Akidau:2013} frameworks by Google, finally, introduce a number of seminal features such as the ``sessionization'' of streaming events that belong to a same operational unit (such as a user session) and a ``watermarking'' technique to handle out-of-order events that fall into a same sliding window.

\begin{figure*}[t]
\centering
\centerline{\includegraphics[width=1\textwidth]{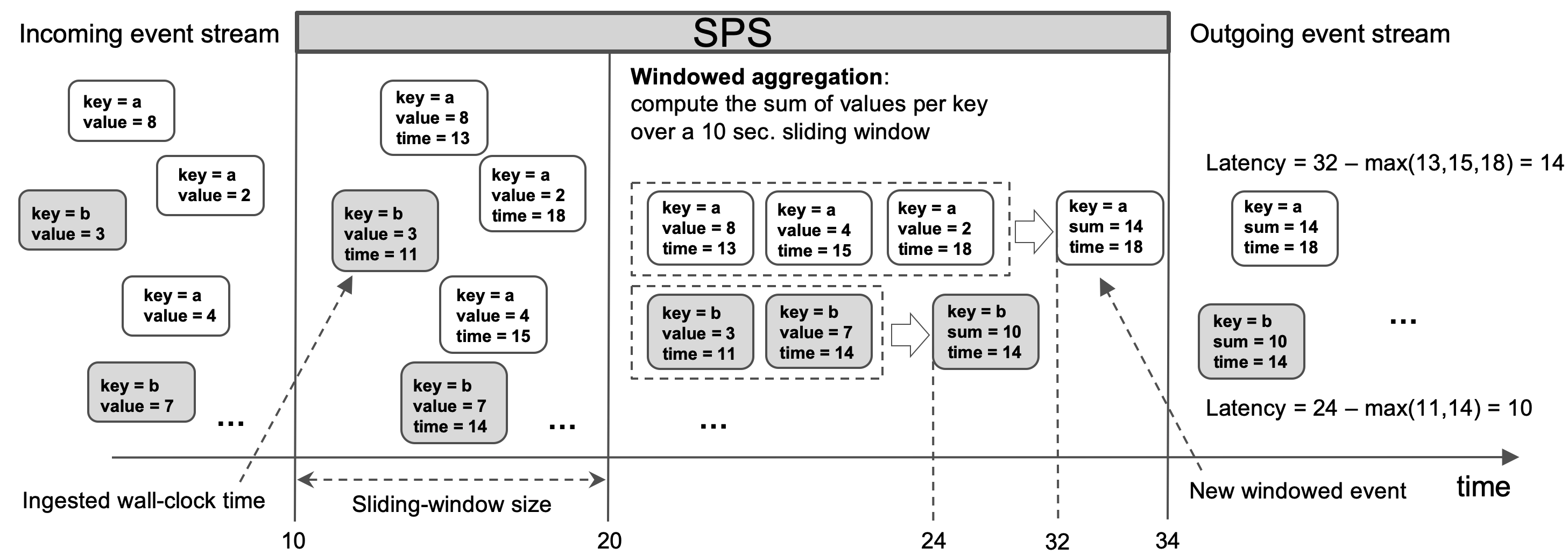}}
\caption{Event-time latency calculation of windowed events}
\label{fig:event-time-latency}
\end{figure*}

\subsection{HPC Resource Allocation}
\label{slurm}
User access and resource management in an HPC cluster usually are facilitated by a workload manager like Moab, Torque, OpenStack or SLURM. In the following, we briefly review the key concepts of SLURM, which is a default open-source solution deployed on many clusters. 

SLURM\footnote{\url{https://slurm.schedmd.com}} is a workload and task management middle-layer for Unix/Linux-based cluster deployments. SLURM users primarily issue jobs via the {\tt salloc}, {\tt sbatch} and {\tt srun} commands. The former two commands create a \emph{resource allocation}, while the latter performs the actual \emph{task allocation} to then run a job in multiple, distributed tasks.  
Typically, {\tt salloc} allocates the requested resources for just a single job, while {\tt sbatch} is used to submit an entire batch of jobs (e.g., a shell script of jobs) to be executed under the same resource allocation.  
The key parameters (amongst various others) used to control both {\tt salloc}/{\tt sbatch} and {\tt srun} are summarized below:
\squishlist
\item {\tt -N}: number of compute nodes
\item {\tt -n}: number of tasks per node
\item {\tt -c}: number of CPUs per task (for multithreading)
\item {\tt -O}: share (i.e., ``overcommit'') CPUs among tasks
\squishend
By default, {\tt  srun} inherits the pertinent arguments of {\tt salloc} and/or {\tt sbatch} and then runs a job in as many tasks as specified by the {\tt -n} argument. This default behavior of {\tt srun} may be overridden by the {\tt -O} option, where the {\tt -n} parameter of {\tt srun} may be greater than given by the actual resource allocation. This allows more tasks to be launched on a single allocation, thus sharing their CPUs equally, which is particularly effective when tasks are internally multithreaded (as it is the case for AIR).

By controlling the above parameters, one may thus gradually move from a centralized, single-threaded execution (e.g., using {\tt srun -N1 -n1 -c1} with a single node, task and CPU) to a distributed and multithreaded execution (e.g., using {\tt srun -N2 -n4 -c4} with 2 nodes, running 2$\times$4=8 tasks and utilizing 4 CPUs per task). 
In ``overcommit'' mode (e.g., using {\tt sbatch -N2 -n4} and then {\tt srun -n16 -O}) multiple tasks (2 per node in this case) execute concurrently on a single resource allocation. SLURM thus allows for a very fine-grained configuration of user sessions and, once granted, also guarantees exclusive access to the requested resources, which makes it an ideal testbed for systematic scale-out tests. 

\subsection{Performance Metrics}
Stream Processing Systems (SPSs) are typically evaluated based on two metrics, namely {\em throughput} and {\em latency} \cite{DSPBM18}. 

\smallskip\noindent\textbf{Event-Time.} 
Two notions of {\em timestamps} are commonly used in SPSs: {\em event-time} and {\em processing-time}. Since we rely on integrated data generators (rather than on an external message brokers such as Kafka), we resort to considering only one notion of timestamps in the remainder of this paper, namely the wall-clock time at which an event is first ingested into the SPS, and we refer to this as the {\em event-time} of the respective event that has been ingested. Using this notion of the event-time thus excludes delays that may arise due to the usage of external data sources or intermediate message brokers. Conversely, the {\em event-time of a windowed event} is defined as the maximum event-time of all events that contributed to that windowed event.

\smallskip
\noindent\textbf{Latency.} The \emph{event-time latency of a windowed event} is defined as the difference between the wall-clock time at which the output event is released by the operator and its event-time. Figure~\ref{fig:event-time-latency} shows the event-time latency calculation of a windowed aggregation, where the input is a set of key-value pairs which each are assigned a timestamp when they are first ingested into the SPS. During the windowed aggregation phase, the sum of the values corresponding to a specific key within a sliding window of 10 seconds is calculated. Once all values corresponding to a key within this window have completed the aggregation, their result is released as a new windowed event whose timestamp is set to the maximum event-time of all events that contributed to it.

\smallskip
\smallskip\noindent\textbf{Throughput.} The {\em throughput} of an SPS is defined as the number of events processed by the system per time unit (typically one second). Thus, to measure the performance of an SPS, it is key to determine the maximum throughput it can handle without exhibiting an undue increase in the queueing (or even dropping) of messages which may in turn lead to a substantial increase in latencies (or even incorrect results)---an effect referred to as ``backpressure'' in~\cite{Kulkarni}. This maximum throughput value then is called the \emph{sustainable throughput} for that particular deployment~\cite{Imai} (see Section~\ref{sec:throughput} for how we determined this value for the systems we compare in the experiments). 

\smallskip
\smallskip\noindent\textbf{Clock Synchronization.} Accurately measuring and comparing event-times in a distributed setting requires a careful synchronization among the local clocks of all compute nodes. We use the {\tt MPI\_Wtime} method of MPI to obtain the wall-clock times of the local compute nodes, which are synchronized by an NTP server with a guaranteed deviation of less than 1 ms.

\section{AIR Architecture}
\label{sec:airarch}
\begin{figure}[!htbp]
\centerline{\includegraphics[width=0.48\textwidth]{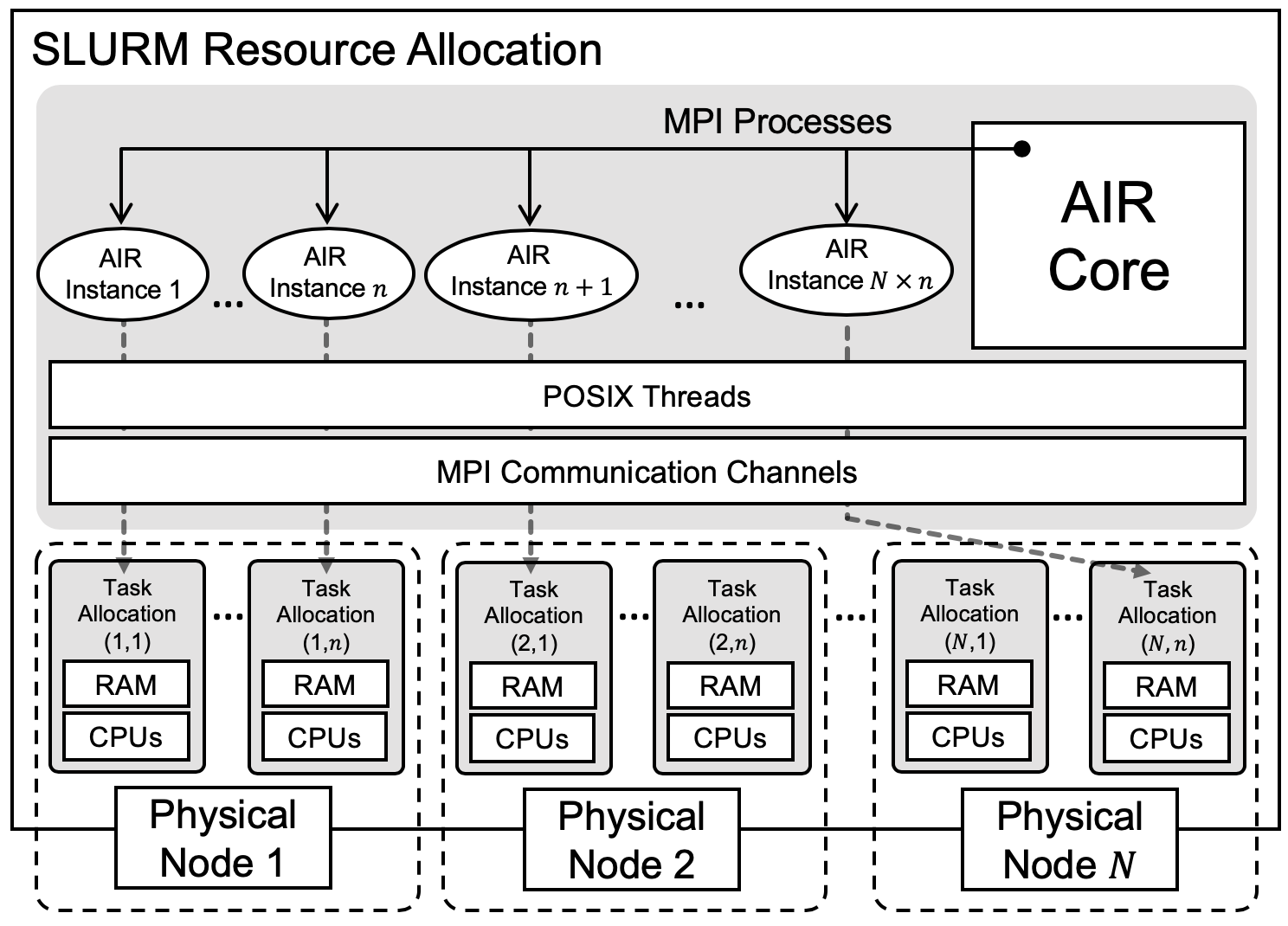}}
\caption{Overview of AIR and its HPC deployment}
\label{arcdgm2}
\end{figure}

The core of AIR is implemented in ANSI C++ using only the Message Passing Interface (MPI) and POSIX Threads (pthreads) APIs, two long-standing libraries for inter-process communication and multithreading, respectively, as additional libraries.  AIR can thus be deployed on any workload manger with a pre-installed MPI package and requires no further dependencies. As shown in Figure~\ref{arcdgm2}, one AIR instance directly corresponds to one task obtained from the SLURM resource allocation. Parallelism is achieved by (1) allocating multiple worker nodes in a cluster, (2) assigning multiple tasks per node, and (3) assigning multiple CPUs to each task. While the former two are facilitated by initializing a separate MPI process (and corresponding AIR instance) for each task, the latter is facilitated via the pthreads library and a respective multithreaded implementation of AIR's dataflow operators. Once instantiated, the AIR instances run in a completely autonomous manner and thereby do not need to rely on a centralized master node that controls the worker nodes. 

\subsection{Physical Architecture}
\label{sec:physicalarch}

\subsubsection{Cluster Deployment (SLURM)}
To use AIR, developers write a dataflow program that implements the high-level control flow of their application in C++, thereby inheriting the features of the core classes (such as {\tt Dataflow} and {\tt Vertex}, described in detail in Section~\ref{sec:logicalarch}) of AIR. Based on a given SLURM resource allocation, $\mathtt{N} \times \mathtt{n}$ separate AIR instances are launched via {\tt srun}, where $\mathtt{N}$ is the number of allocated nodes and $\mathtt{n}$ is the number of allocated tasks per node.
Internally, AIR thus seamlessly integrates with the SLURM environment via the MPI APIs, whose principal parameter, namely the {\em worldSize}, is directly inherited as the number $\mathtt{N} \times \mathtt{n}$ of AIR instances that have been initialized via SLURM.
As shown in Figure~\ref{arcdgm2}, SLURM has performed a number of task allocations such that each of the $\mathtt{N}$ physical compute node runs $\mathtt{n}$ AIR instances with exclusive access to $\mathtt{c}$ CPU cores each. As outlined in the previous section, the ``overcommit'' ({\tt -O}) option may further be utilized to also grant shared access to these resources among multiple AIR instances.

\subsubsection{Inter-Process Communication (MPI)}
\label{sec:comm}
AIR by default uses the MPICH\footnote{\url{https://www.mpich.org}} C/C++ libraries but can also dynamically be linked with other MPI libraries such as OpenMPI or MVAPICH. Via MPI, each AIR instance has a holistic view of the entire set of allocated tasks and can thus reroute messages from itself to any other instance without depending on a centralized master node. 

\smallskip
\noindent\textbf{Initialization.} Two static MPI functions, {\tt MPI\_Comm\_size} and {\tt MPI\_Comm\_rank}, return the {\em worldSize} (i.e., the overall number of available tasks across all compute nodes) and the {\em rank} (i.e., the unique index of the current task), respectively. As shown in Algorithm~\ref{alg:dataflow}, both parameters are used to initialize a {\tt Dataflow} class which represents the main entry point to the AIR core libraries. The {\em worldSize} conforms to the combination of {\tt -N} and {\tt -n} parameters of SLURM specified when launching the AIR binary via {\tt srun} (e.g., issuing {\tt srun -N2 -n4 -c4 ./AIR} would result in a {\em worldSize} of 8 where each task has exclusive access to 4 CPU cores).

\smallskip
\noindent\textbf{Communication.} MPI provides two {\em blocking methods}, {\tt MPI \_Send} and {\tt MPI\_Recv}, for sending and receiving binary message blocks among ranks in the network. Both allow for specifying the target and source ranks as well as a unique {\em tag} that needs to be matched among a sent and received message. We remark that MPI provides also two {\em non-blocking counterparts}, {\tt MPI\_Isend} and {\tt MPI\_Irecv}, which provide their own internal threading mechanism to asynchronously send and receive messages. For AIR, we however do not rely on the latter two, since we wrap {\tt MPI\_Send} and {\tt MPI\_Recv} into our own multithreaded communication protocol which allows us to precisely control the queuing and processing of incoming and outgoing messages. 

\smallskip
\noindent\textbf{Channels.} All pairs of ranks (including pairs of same ranks) may communicate among each other by exchanging messages. A basic {\em channel} is a pair of $\langle \mathit{sourceRank}, \mathit{targetRank} \rangle$ indices, in which each channel has a unique sender and receiver. 
Channels are not directly available in MPI, but a similar form of determinism in the communication is achieved by encoding the channel information into the {\em message tag} used by the {\tt MPI\_Send} and {\tt MPI\_Recv} calls.
In the case of a DAG-structured dataflow model, in which multiple operators run both in a distributed and parallel manner per rank, the notion of a channel needs to be extended by including also the {\em source} and {\em target ids} of the operators that exchange messages among each other. We therefore define the {\em 32-bit message tag} associated with each incoming and outgoing message as
$\mathit{sourceRank} \ll 24 +  \mathit{sourceOp} \ll 16 + \mathit{targetRank} \ll 8 + \mathit{targetOp}$.


\subsubsection{Intra-Process Multithreading (pthreads)}
AIR employs multithreading to control (i) the queuing of incoming messages, (ii) the processing of the received messages, and (iii) the sending of outgoing messages at each dataflow operator and communication channel individually (see also Algorithm~\ref{alg:vertex} of the {\tt Vertex} class). Specifically, {\tt pthread\_create} and {\tt pthread\_join} are used to invoke and join (i.e., provide a synchronized termination) of these threads. Moreover, for internal thread synchronization, we rely on the {\tt pthread\_mutex\_lock}, {\tt pthread\-\_cond\_wait} and {\tt pthread\_ mutex\_unlock} methods to take a lock on a mutex variable, wait for a signal being sent to a conditional variable, and release the lock on a mutex variable, respectively. The latter three methods are used to implement {\em producer-consumer patterns} between (i-ii) the thread receiving the incoming messages and the thread-processing these messages, and between (ii-iii) the processing thread and the thread sending the outgoing messages at each channel.

\subsection{Logical Architecture}
\label{sec:logicalarch}

\subsubsection{Dataflow Programs}
AIR adopts the notion of {\em dataflow programs}~\cite{Johnstondf} in which a DAG-structured execution plan of {\em logical data transformations} is compiled into a physical execution plan of stream-processing operators. Formally, a dataflow is a {\em multi-rooted, directed acyclic graph} (DAG) in which the vertices represent the transformations we wish to execute against one or more incoming data streams with respect to the topology of the DAG. The roots of the DAG are so-called {\em sources}, which implement C/C++ interfaces to the streaming data sources (e.g., to message brokers such as Kafka or to other web services via CURL), large files, or custom data generators. Conversely, the leaves of the DAG are so-called {\em sinks}, which usually write the aggregated results of the dataflow transformations into a persistent storage (such as Redis or output files), or which in turn provide streaming access to other SDPSs. Internal vertices of the DAG represent either unary (such as {\tt Map}, {\tt Reduce}) or binary (such as  {\tt Join}) transformations that each connect to either one or two input streams and produce one or more output streams. The logical architecture of AIR is captured via the pseudocode provided in Algorithms~\ref{alg:dataflow}--\ref{alg:reduce}, which we describe in detail in the following steps.

\smallskip\noindent\textbf{Dataflow.} The {\tt Dataflow} class depicted in Algorithm~\ref{alg:dataflow} serves as the main entry point to create a number of concurrent AIR instances. After a SLURM session has been established with the desired number of nodes, tasks per node, and CPU cores per task, one instance of the {\tt Dataflow} class is created per SLURM task. This is indicated by the {\tt main} method in Algorithm~\ref{alg:dataflow}, which is merely parameterized by the number of desired AIR tasks. Next, each {\tt Dataflow} instance reads the current {\em worldSize} and {\em rank} parameters from the MPI environment and thereby runs each AIR task in a shared-nothing manner. Each {\tt Dataflow} instance has access to the topology of the entire dataflow program by its {\tt vertices} internal vector of interlinked dataflow operators (subclasses of {\tt Vertex}). The {\tt streamProcess} method then in parallel invokes the three types of threads by calling the {\tt listeningThread}, {\tt pro\-cessingThread} and {\tt sendingThread} thread entry points---one per vertex instance and communication channel.
\begin{algorithm}[htb]
\scriptsize
\vspace{1mm}
\begin{verbatim}

void main(int #tasks) {

  // Initialize one instance of Dataflow per task
  for i from 1 to #tasks in parallel do { 
   Dataflow dataflow = new Dataflow();
   dataflow.streamProcess(); }
} 

class Dataflow {
  
   // Basic MPI parameters
  int worldSize, rank; 
  
  // Vector holding the dataflow operators
  vector<Vertex> vertices; 
  
  // Initialize MPI environment at current task
  Dataflow() { 
    MPI_Init_(); 
    MPI_Comm_size(worldSize); 
    MPI_Comm_rank(rank); 
  }

  // Main entry point to invoke stream processing 
  void streamProcess() {
    
    // Invoke listener, processing & sender threads per vertex
    for each vertex in vertices in parallel do {
    	vertex.startThreads(); }
  }
}
\end{verbatim}
\vspace{1mm}
\caption{Dataflow pseudocode of AIR}
\label{alg:dataflow}
\end{algorithm}

\smallskip\noindent\textbf{Vertex.} The {\tt Vertex} class depicted in Algorithm~\ref{alg:vertex} implements the operators involved in the dataflow program and thereby also facilitates the communication among these operators in a distributed deployment of AIR. According to the topology of the dataflow DAG, each vertex is initialized with 
$\mathit{indegree} \times \mathit{worldSize}$
many {\em incoming channels} (i.e., one per incoming vertex and rank) and
$\mathit{outdegree} \times \mathit{worldSize}$
many {\em outgoing channels} (i.e., one per outgoing vertex and rank). The respective {\tt inChannels} and {\tt outChannels} vectors thus represent mechanisms both for thread synchronization and the queueing of incoming and outgoing messages. The actual synchronization and communication logic is inherited and thus identical for all subclasses of {\tt Vertex}. Correspondingly, one listening and processing thread is created per incoming channel, while one sending thread is created per outgoing channel (as indicated by the respective methods of {\tt Vertex}). The threads and respective message queues are synchronized via producer-consumer patterns to decouple the message passing from their processing as much as possible.

\begin{algorithm}[htb]
\scriptsize
\vspace{1mm}
\begin{verbatim}

class Vertex {
 
  // Alive flag for threads  
  bool ALIVE = false;
  
  // Channels for thread synchronization & MPI protocol
  vector<Channel> inChannels;
  vector<Channel> outChannels;
  
  void startThreads() {
    
    ALIVE = true;
    
    // Invoke listener & processing thread per incoming channel
    for each inChannel in inChannels in parallel do {
      pthread_create( listeningThread(inChannel) );
      pthread_create( processingThread(inChannel) ); }
  
    // Invoke sender thread per outgoing channel    
    for each outChannel in outChannels in parallel do {    
      pthread_create( sendingThread(outChannel) ); } 
  }
  
  void listeningThread(inChannel) {
     while (ALIVE) {
       inMessage = new Message();
       MPI_Recv(inMessage);
       inChannel.push_back(inMessage); 
       inChannel.notify(); }
  }

  void processingThread(inChannel) {
    while (ALIVE) {
      inChannel.cond_wait();
      inMessage = inChannel.pop(); 
      
      // Vector of outgoing messages, one per outgoing channel
      vector<Message> outMessages;
      
      // Actual process method will be overwritten by subclass
      process(inMessage, outMessages); 
      
      // Asynchronous iterative routing of outgoing messages
      for i from 1 to outChannels.size in parallel do {
        outChannel[i].push_back(outMessage[i]);
        outChannel[i].notify(); }
      
      outMessages.clear();
      delete inMessage; }
  }

  void sendingThread(outChannel) {
    while (ALIVE) {
      outChannel.cond_wait();
      outMessage = outChannel.pop(); 
      MPI_Send(outMessage);
      delete outMessage; }
  }
}
\end{verbatim}
\vspace{1mm}
\caption{Vertex pseudocode of AIR}
\label{alg:vertex}
\end{algorithm}

\smallskip
\noindent\textbf{Listening Thread.} 
The {\em listening thread} of each incoming channel uses the blocking method {\tt MPI\_Recv} and fills the {\tt inChannel} message queue with one incoming message per iteration of the embracing {\tt while} loop. The received message must match the {\em message tag} computed from the channel information that each thread processes (see Section~\ref{sec:comm}). Upon having received a message, the listening thread notifies the processing thread at the same channel and immediately turns into the next iteration to receive a new message.

\smallskip
\noindent\textbf{Processing Thread.} 
Once notified by the listening thread, the {\em processing thread} pops each incoming message from its assigned {\tt inChannel} and calls the {\tt process} stub, which must be overwritten by each subclass of {\tt Vertex} with the intended logic of that subclass (see Section~\ref{sec:dataflowops} for various operator specifications).
We remark that {\tt process} has access to both the incoming message as well as to an entire vector {\tt outMessages} of outgoing messages. In the final step of each processing iteration, the new outgoing messages are placed into the outgoing message queues associated with the {\tt outChannels} vector, and the respective sending threads are notified.

\smallskip
\noindent\textbf{Sending Thread.} 
Once notified by the processing thread, the {\em sending thread} pops each message from its assigned {\tt outChannel} queue and uses the blocking method {\tt MPI\_Send} to send the outgoing message to the rank associated with this channel. After the message has been successfully sent, the message can also safely be deleted by this thread.

\smallskip
All threads remain alive until some external termination signal is sent (also via messages), which indicates that the execution of AIR should be completed.

\subsection{Asynchronous Iterative Routing} 
\label{sec:air}
As outlined in the previous subsection, the processing thread at each incoming channel has access to an entire vector of outgoing message queues---one for each outgoing channel. Each processing thread therefore is able to ``cross'' channels and thereby dynamically reroute messages to all ranks, including itself, in the MPI environment (or  tasks in SLURM). Since all threads asynchronously receive, process and send messages, we refer to this technique as ``Asynchronous Iterative Routing'' (AIR). The exact routing strategy is both {\em workload-} and {\em application-dependent}; a simple {\tt Map} transformation may simply reroute data elements according to a random sharding technique such as $\mathit{value} \,\%\, \mathit{worldSize}$, while a {\tt Split} operation will have to separate the data values and thus route each selected value to a different branch in the dataflow DAG (compare to Figure~\ref{fig:extendedYSB} from the Introduction). A {\tt Join} operation, finally, needs to take the respective join keys into account to make sure that tuples with the same join keys are indeed routed to the same ranks. The exact routing strategy thus is part of the {\tt process} method implemented by each subclass of {\tt Vertex}.

\subsection{Pipelining} 
\label{sec:pipe}
Once one or more incoming data streams have been evenly routed to all ranks, most data transformation operators such as {\tt Map} or {\tt StaticJoin} do not actually need to perform rerouting to guarantee a correct execution of the dataflow program. {\em Pipelining} thus is a global flag associated with each operator in the dataflow DAG. At each rank, this flag enables an operator to directly place an outgoing message from its {\tt outChannel} outgoing message queue into the {\tt inChannel} incoming message queue of the next operator in the DAG's topology. Thus, pipelining is able to circumvent MPI communication in order to directly exchange messages between operators within a same rank, which may substantially accelerate performance when network communication is not needed to actually redistribute data.

\begin{algorithm}[htb]
\scriptsize
\vspace{1mm}
\begin{verbatim}

class Map : Vertex {

  // Function associated with this vertex subclass
  Function function;
  
  // Initialize Map with actual function to calculate
  Map (function) {
    this->function = function;
  }
  
  void process(inMessage, outMessages) {
    
    // Deserialized partition for current message
    Partition partition = deserialize(inMessage);
    
    // Transform all events inside this partition
    for each event in partition do {
       event.value = function.calculate(event.value); }
       
    // Serialize partition & send messages
    outMessages = serialize(partition); 
  }
}
\end{verbatim}
\vspace{1mm}
\caption{Map pseudocode of AIR}
\label{alg:map}
\end{algorithm}

\begin{algorithm}[htb]
\scriptsize
\vspace{1mm}
\begin{verbatim}

class Reduce : Vertex {

  // Aggregate function associated with this vertex subclass
  AggregateFunction function;
  
  // Map from event-times to partitions
  Map windowedPartitions;
  
  // Window size associated with this operator
  int windowSize; 
  
  // Initialize Map with actual function to calculate
  Reduce (function, windowSize) {
    this->function = function;
    this->windowSize = windowSize;
    this->windowedPartitions = new Map();
  }
  
  void process(inMessage, outMessages) {
  
    // Deserialized partition for current message
    Partition partition = deserialize(inMessage);
    
    // Get windowed partition of each event or create new partition
    for each event in partition do {
       wId = event.timestamp / windowSize;
       wPartition = windowedPartitions.getOrElse(wId, 
           new Partition());
       function.combine(wPartition, event.value); }
    
    // Serialize & send each complete windowed partition
    for each wPartition in windowedPartitions do {
       if wPartition.isComplete
          outMessages = serialize(partition); }
   }
}
\end{verbatim}
\vspace{1mm}
\caption{Reduce pseudocode of AIR}
\label{alg:reduce}
\end{algorithm}

\section{Dataflow Operators}
\label{sec:dataflowops}

\subsection{Data-Transformation Operators}
Data-transformation operators, such as {\tt Map}, {\tt Filter}, {\tt Split}, {\tt StaticJoin}, etc., take one event at a time as input and perform an associated transformation function on each of these events individually. These one-to-one transformations thus process events independently of each other and therefore do not require synchronization. Also timestamps can be ignored inside the logic of these operators and are merely forwarded to the sliding-window operators. Therefore, we refer to these operators as \emph{stateless}, i.e., they do not need to maintain any context information about the sliding window to which each of the processed events belongs.

\begin{figure}[!h]
\centerline{\includegraphics[width=0.40\textwidth]{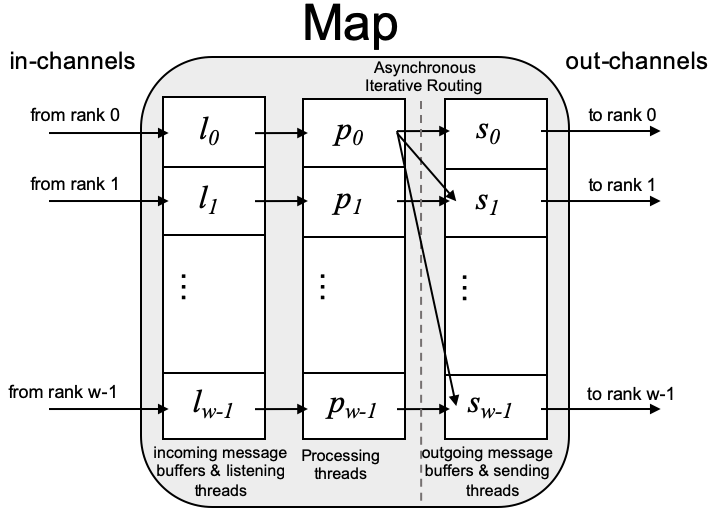}}
\caption{Map with an in- and outdegree of 1}
\label{map}
\end{figure}
\smallskip
\noindent\textbf{Map.} The {\tt Map} class is a subclass of {\tt Vertex} and therefore inherits all of the latter's mechanisms for communication via channels and for thread synchronization. It is initialized with a dedicated {\tt function} that implements the actual calculation associated with {\tt Map}. The pseudocode for {\tt Map} is given in Algorithm~\ref{alg:map}. Since {\tt Map} is a one-to-one transformation (e.g., applying ``times-two'' as transformation function), each element of a deserialized data partition that is obtained from an incoming message is applied to that function at a time (or in parallel). The result of each calculation is in turn serialized back into the outgoing messages (one per outgoing channel), which are then queued and sent by the respective sender threads in an asynchronous manner---the key idea for what we refer to ``Asynchronous Iterative Routing'' (see Section~\ref{sec:air}). A typical {\tt Map} operator takes the value associated with an event and applies a random sharding strategy by serializing the event into the outgoing message with index $\mathit{value} \,\%\, \mathit{worldSize}$. If pipelining is enabled, however, all values are directly serialized into the outgoing message with index $\mathit{rank}$ at the current rank and directly copied into the incoming message queue of the subsequent operator according to the topology of the dataflow program.
Figure~\ref{map} depicts the internal architecture of the {\tt Map} operator, where $l_i$, $p_i$ and $s_i$ indicate listening, processing and sending threads, respectively ($w$ corresponds to the $\mathit{worldSize}$ parameter)

\smallskip
\noindent\textbf{Filter.} This is a special case of {\tt Map}, where the {\tt function} associated with the operator takes a deserialized data element  as input and returns a Boolean value. Based on the Boolean value returned, the processing thread determines if that data element should be serialized or not into the outgoing messages. {\tt Filter} behaves analogously to {\tt Map} in terms of routing and pipelining.

\begin{figure}[!h]
\centerline{\includegraphics[width=0.40\textwidth]{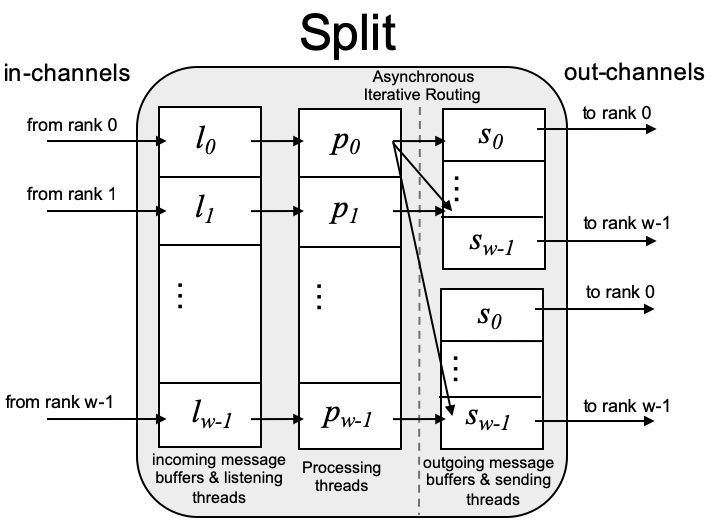}}
\caption{Split with an outdegree of 2}
\label{split}
\end{figure}

\smallskip
\noindent\textbf{Split.} Also this operator is similar to {\tt Map}. However, here events are distributed according to two or more splitting conditions. For routing, the index of an event consisting of a $\langle\mathit{key}, \mathit{value}\rangle$ pair in the outgoing message vector is computed as $\mathit{index}(\mathit{key}) \times \mathit{worldSize} + \mathit{value} \,\%\, \mathit{worldSize}$, where $\mathit{index}(\mathit{key})$ denotes the index of a splitting condition among the operator's successors in the topology of the dataflow program. The architecture of the {\tt Split} operator with an outdegree of 2 is depicted in Figure~\ref{split}.

\smallskip
\noindent\textbf{StaticJoin.} This operator joins each incoming event with a static map (e.g., read from an external file that is shared with all worker nodes) which is kept in the main-memory of all worker nodes. The augmented events are then serialized into the outgoing messages, where an analogous strategy regarding routing and pipelining as for {\tt Map} is used.

\subsection{Sliding-Window Operators}

Sliding-window operators, such as {\tt Reduce}, {\tt Aggregation}, {\tt Join}, etc., group multiple events, namely those that fall into the same sliding window according to their timestamps, in order to perform their associated aggregation or join operation. Since, especially in an ASP architecture, sliding windows cannot be assumed to be aligned with the batch or message sizes, these operators require synchronization by the respective window ids. For the same reason, we refer to these operators as \emph{stateful}, i.e., 
they need to maintain an in-memory data structure (usually a hash map) across all processing threads inside each operator, which stores intermediate results that fall into a same window. Consequently, multithreaded access to this data structure needs to be synchronized via a mutex to guarantee correct results.


\smallskip\noindent\textbf{Reduce.} The {\tt Reduce} class is a subclass of {\tt Vertex} and therefore inherits all of the latter's mechanisms for communication via channels and for thread synchronization. It is initialized with an {\tt AggregateFunction} that implements the actual calculation associated with {\tt Reduce}. The pseudocode for {\tt Reduce} is given in Algorithm~\ref{alg:reduce}. Since {\tt Reduce} is a many-to-one aggregation (e.g., applying ``sum'' as aggregation), multiple deserialized elements obtained from the incoming messages need to be merged into a single result. Moreover, if elements of a sliding window are spread across multiple messages (which is typically the case in an asynchronous setting), {\tt Reduce} may only release the results of each sliding window once it knows that all elements of that window have been aggregated. The result of each window is in turn serialized into one of the outgoing messages (one per outgoing channel). The routing and pipelining of outgoing events is again analogous to {\tt Map}.

\smallskip\noindent\textbf{Aggregation.} This operator is similar to {\tt Reduce} except that events consist of $\langle\mathit{key}, \mathit{value}\rangle$ pairs, and the aggregate function is performed to all values that fall under the same $\mathit{key}$. {\tt Aggregation} uses an analogous strategy regarding routing and pipelining as {\tt Reduce}, however using keys instead of values.

\begin{figure}[!h]
\centerline{\includegraphics[width=0.40\textwidth]{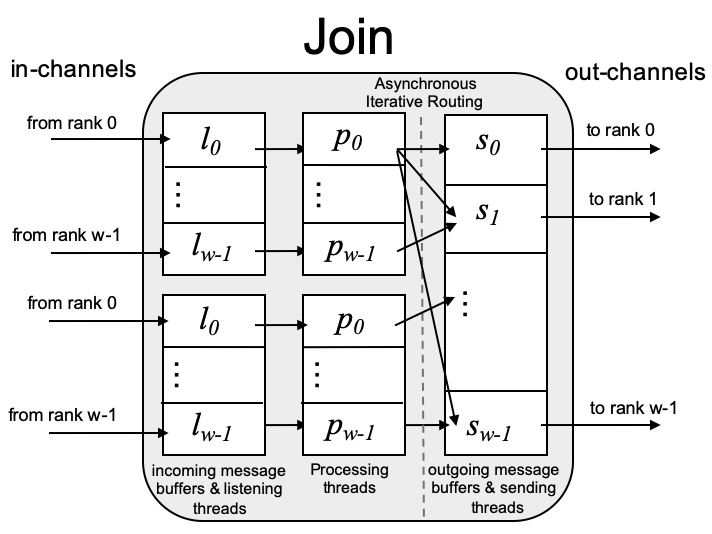}}
\caption{Join with an indegree of 2}
\label{join}
\end{figure}

\smallskip
\noindent\textbf{Join.} The {\tt Join} operator implements an equi-join among two streams of incoming events with compatible $\langle\mathit{key}, \mathit{value}\rangle$ pairs. An analogous strategy regarding routing and pipelining as for {\tt Reduce} and {\tt Aggregation} is used also here. Figure~\ref{join} depicts the internal architecture of the {\tt Join} operator with an indegree of 2.

\subsection{Local vs. Global Aggregations}
For stateful sliding-window operators, such as {\tt Reduce}, {\tt Aggre\-gation} or {\tt Join}, it is generally beneficial to first perform a {\em local pre-aggregation} among events consisting of $\langle\mathit{key}, \mathit{value}\rangle$ pairs, during which the events are first rerouted among all ranks according to their keys in order to improve parallelism. In a subsequent {\em global aggregation} step, the pre-aggregated events are then once more rerouted based on their window ids (calculated from their timestamps) to fully aggregate and synchronize these events with respect to the sliding windows they belong to. For {\em commutative} and {\em associative} aggregation functions, such as {\em sum}, this form of pre-aggregation does not affect the results but helps to increase performance. Internally, the pre- and full-aggregation steps are thus implemented as two stateful operators, which each employ two levels of nested hash maps. In the outer map, the keys are the window ids; while in the inner maps, the keys are the $\mathit{key}$ entries of the respective events. This form of nested hashing is outlined in the {\tt process} method of {\tt Reduce} (Algorithm~\ref{alg:reduce}), where first a {\tt wId} is computed from the timestamp of an event, whereupon the event is then merged and aggregated into the respective partition of events that fall into this window.
%
%

\section{Experiments}
\label{sec:exp}

The experiments we run with AIR in comparison to Spark and Flink are three-fold.  First, we begin by determining the {\em sustainable throughput} under the given HPC setup for all systems under investigation: AIR, Spark and Flink, which requires many repeated runs at different throughput levels. For AIR, we need to repeat these runs for each choice of the {\tt -N}, {\tt -n} and {\tt -c} parameters, while for Spark and Flink the best setup is usually given by allocating all resources for a given number of nodes (thus only the {\tt -N} parameter is relevant for these systems). Second, based on the sustainable throughput values we obtained from the first step, we perform a systematic {\em scale-out test} of AIR on our HPC cluster under various task-allocation parameters. Third, we compare the sustainable throughput rates of the three systems under three different use-cases, namely (1) a {\em Simple Windowed-Aggregation} (SWA), (2) the original version of the {\em Yahoo Streaming Benchmark} (YSB)~\cite{ysb}, and (3) the extended version of YSB (called YSB*) as it is depicted in Figure~\ref{fig:extendedYSB}.

\subsection{General Setup \& Benchmarks}
Each node of our HPC cluster is equipped with two 2.6 GHz Intel Xeon Gold 6132 CPUs, 28 cores and 128GB RAM. All nodes run CentOS Linux (v7) and are managed by SLURM (v19.05.3). 
The network bandwidth is 100Gb/s (Infiniband EDR). 
The benchmark implementations of AIR are complied with GCC (v6.4.0) with -O3 optimization. The only additional libraries are Intel MPI (v18.0.1) and the default POSIX Thread API of GCC. We use Apache Flink (v1.6.1) and Apache Spark Streaming (v2.3.2) as our baseline DSPSs. We have disabled the checkpointing options in these systems to minimize their additional overheads. The Scala (v2.11.12) implementations of the YSB benchmarks run on top of JDK v1.8.0. All scripts used for Spark and Flink are also available via our GitLab repository\footnote{https://gitlab.uni.lu/vvenugopal/streambenchmarks}.

\subsubsection{Simple Windowed-Aggregation (SWA)} 
\label{sec:swa}
The first benchmark consists of an in-memory data generator and a single windowed aggregation. We reuse the same data generator as the one implemented for YSB (see next subsection) to generate variable length events (on an average 136 bytes) at various throughput rates. The windowed aggregation then groups events over a 10-second sliding window and counts the number of events produced by the generator. 
For Spark and Flink, we rely on their default grouping and aggregation operators (such as {\tt groupBy(.)}, {\tt keyBy(.)}, {\tt agg(.)}) for streams. 
The data generator is directly implemented in the native runtime of all systems (Scala in the case of Spark and Flink). Just like any other operator, it runs in a distributed manner and generates the input events in multiple tasks in parallel. In AIR, this dataflow contains 3 operators, namely {\tt Generator}, {\tt Aggregation} plus a final sink node that writes the results produced by all ranks into a single file. 

\subsubsection{Yahoo Streaming Benchmark (YSB)}
\label{sec:ysb}
The original Yahoo Streaming Benchmark~\cite{ysb} simulates a simple advertise\-ment-based analytics pipeline, where a stream of ad events is consumed from a message broker, and the application computes a sliding-window aggregation of ads being mapped uniformly across 100 distinct campaigns that have been "viewed" by a user. In recent streaming benchmark designs (as proposed by the Flink and Spark communities)~\cite{DSPBM18, DataBrick}, to accurately measure the performance of the DSPSs, distributed in-memory data generators are utilized instead of external message brokers such as Apache Kafka~\cite{kreps2011kafka}. Therefore, we employ the modified YSB implementation given in~\cite{DataBrickGit} to avoid potential bottlenecks due to external message brokers like Kafka or Redis for storage. This guarantees that the data-generation rate is able to actually keep up with the data-ingestion rate of the streaming system up to very high throughputs and at the same time eliminates latencies due to these external components. In AIR, the YSB dataflow contains 5 operators, namely {\tt Generator}, {\tt Filter}, {\tt StaticJoin}, {\tt Aggregation}, and again the sink node to store the results.

\subsubsection{Yahoo Streaming Benchmark (YSB*)}
\label{sec:ysb*}
As mentioned in the previous subsection, the original YSB follows a simple pipeline, where we have only one windowed aggregation, while the join from ads to campaigns remains a static mapping based on an in-memory table that is shared among all worker nodes. To make the use-case more challenging (without affecting the basic benchmark setup), we modified the YSB dataflow as shown in Figure~\ref{fig:extendedYSB}.  
In this modified form of YSB, called YSB*, events are first split into two streams: one corresponding to ``Click'' events and another one corresponding to ``View'' events. The two event streams then each are transformed in a similar manner as in YSB. The counts of view- and click-events per campaign are calculated by two independent windowed aggregations. Further, we introduce a third windowed operator, which joins the corresponding events produced by the previous aggregations to finally determine the ratio of click- and view-events per campaign.

In this extended use-case, the major challenge consists of performing two windowed operations in parallel and to then join their results by a third windowed operation: first compute the counts of click- and view-events individually, then compute their ratios. The streaming API of Flink has a {\tt Join} function for combining two streams and also supports multiple window-based operations in a sequence. However, Spark (as of its recent 2.4 version) does not support multiple window-based operations. Therefore, we have utilized the multiple simultaneous aggregation\footnote{Performing multiple aggregation operations simultaneously on a single time-window based group than performing those aggregations on multiple consecutive time-window based groups} feature in the structured-streaming API to implement YSB* in Spark~\cite{DataBrick2}. 
In AIR, this use-case is implemented using the 9 operators shown in Figure~\ref{fig:extendedYSB}) (just the final sink node that writes the results from all ranks into a single file is not shown in the figure). Most of these operators have the same functionality as we saw in the YSB use-case already---except for the {\tt Join} which is a new kind of a windowed operator that comes into play here. 


\subsection{Sustainable Throughput}
\label{sec:throughput}
To determine the \emph{sustainable throughput} (ST) for a given cluster deployment, we run each system with a low throughput first and gradually increase the throughput until the system shows back-pressure. Back-pressure in these systems is identified based on the following observations:
\squishlist
\item For Spark and AIR, the system is said to be in back-pressure if a sudden raise in (average) event-time latency is observed while increasing the input throughput beyond a sustainable range. The green lines in Figures \ref{f1} and \ref{f2} depict the ST values for Spark and AIR, respectively.
\item For Flink, it is observed that---unlike in Spark and AIR---the (average) event-time latency does not raise abruptly when the throughput goes beyond the sustainable value. Instead, we see a sudden decline in the count of processed windows. Therefore, in addition to the latency value, the ST is determined by considering the window count as well. Figure~\ref{f3} shows the correlation of latency with the count of processed windows under different throughput values.
\squishend

\smallskip
\noindent
The sustainable throughput is specific to each workload and cluster deployment of a system. We ran the YSB benchmark on $N=2$ nodes for 300 seconds, thus processing 30 sliding windows of 10 seconds each to compute the average latency among the aggregated results. For Spark, while increasing the throughput from 5M to 12M events per second, we can observe a sudden rise in the average latency from 877 ms to 5,939 ms, which steadily keeps increasing from there on. For AIR (using $N=2$, $n=2$ and $c=4$), we made a similar observation when the throughput was around 24M with a latency of 1,423 ms. Flink remains more stable with respect to average latency. However, the number of windows processed suddenly drops to 786 from 2,999 at a throughput of 13.5M events per second, which indicates that data is actually dropped.

\begin{figure*}[t]
\begin{multicols}{3}
\begin{minipage}{.3\textwidth}
\caption{ST for Spark on 2 nodes\label{f1}}
     \begin{tikzpicture}[scale=.9]
     \begin{axis}[
       width=0.42*2.5\textwidth,
       height=0.30*2.5\textwidth,
       axis x line=bottom,
       xmin=0,
       xmax=45,
       xtick={0,10,20,30,40},
        xlabel={throughput (in 10$^6$ events/sec.)},
       xlabel near ticks,
       xticklabel style={/pgf/number format/1000 sep=},
       axis y line=left,
       ymin=0,
       ymax=48672,
       ylabel={avg. latency (ms.)},
       ylabel near ticks
     ]
       \addplot[color=red,mark=x] coordinates {
	(1,681)
	(5,877)
	(12.5,5938)
	(13,15938)
	(20,42719)
	(30,58672)
	(40,78672)
       };
 	\draw[color=green] (axis cs:12.5,\pgfkeysvalueof{/pgfplots/ymin}) -- (axis cs:12.5,\pgfkeysvalueof{/pgfplots/ymax});
    \end{axis}
   \end{tikzpicture}
\end{minipage}

\begin{minipage}{.3\textwidth}
\caption{ST for AIR on 2 nodes\label{f2}}
 \begin{tikzpicture}[scale=.9]
     \begin{axis}[
       width=0.42*2.5\textwidth,
       height=0.30*2.5\textwidth,
       axis x line=bottom,
       xmin=0,
       xmax=45,
       xtick={0,10,20,30,40},
       xlabel={throughput (in 10$^6$ events/sec.)},
       xlabel near ticks,
       xticklabel style={/pgf/number format/1000 sep=},
       axis y line=left,
       ymin=0,
       ymax=45000,
       ylabel={avg. latency (ms.)},
       ylabel near ticks
     ]
       \addplot[color=red,mark=x] coordinates {
(1,192)
(5,365)
(13,578)
(22,799)
(23,833)
(24,1423)
(25,6496)
(26,15794)
(27,19737)
(28,25990)
(30,35008)
(31,38964)
(32,44304)
       };
 \draw[color=green] (axis cs:24,\pgfkeysvalueof{/pgfplots/ymin}) -- (axis cs:24,\pgfkeysvalueof{/pgfplots/ymax});
     \end{axis}
   \end{tikzpicture}
\end{minipage}

\begin{minipage}{.3\textwidth}
\caption{ST for Flink on 2 nodes\label{f3}}
   \begin{tikzpicture}[scale=.9]
     \begin{axis}[
       width=0.38*2.1\textwidth,
       height=0.35*2.1\textwidth,
       axis x line=bottom,
       xmin=0,
       xmax=160,
     xlabel={throughput (in 10$^6$ events/sec.)},
       xlabel near ticks,
       xticklabel style={/pgf/number format/1000 sep=},
       axis y line*=left,
       ymin=0,
       ymax=9000,
       ylabel={avg. latency (ms.)~\ref{latency}},
       ylabel near ticks
     ]
       \addplot[color=red,mark=x] coordinates {
(1,1681)
(5,1682)
(10,1948)
(13.5,1766)
(20,1632)
(30,3869)
(33,2938)
(40,1719)
(80,2049)
(160,2672)
       };\label{latency}

\draw[color=green] (axis cs:13.5,\pgfkeysvalueof{/pgfplots/ymin}) -- (axis cs:13.5,\pgfkeysvalueof{/pgfplots/ymax});
\end{axis}
\begin{axis}[
width=0.295*1.715\textwidth,
height=0.26*1.715\textwidth,
scale only axis,
xmin=0,
xmax=160,
xlabel near ticks,
xticklabel style={/pgf/number format/1000 sep=},
axis y line=right,
xtick=\empty,
x axis line style={draw opacity=0},
ymin=0,
ymax=4000,
ylabel style = {align=center},
ylabel={windows processed~\ref{wincount}},
]
 \addplot[color=blue,mark=x] coordinates {
(1,2999)
(5,2999)
(10,2999)
(13.5,2892)
(20,786)
(30,786)
(33,786)
(40,300)
(80,238)
(160,120)
       };
\label{wincount}
\end{axis}
\end{tikzpicture}
\end{minipage}
\end{multicols}
\end{figure*}

\definecolor{babyblue}{rgb}{0.54, 0.81, 0.94}
\definecolor{aureolin}{rgb}{0.99, 0.93, 0.0}
\definecolor{bittersweet}{rgb}{1.0, 0.44, 0.37}
\definecolor{cream}{rgb}{1.0, 0.99, 0.82}

\begin{figure*}\caption{ST values (in 10$^6$ events/sec.) of various AIR modes and under different HPC deployments for the 3 use-cases: SWA, YSB and YSB*}
\label{fig:metrics}
\begin{tikzpicture}
\begin{groupplot}[
legend entries={{\color{red}{\tiny a1}},{\color{blue}{\tiny a2}},{\color{black}{\tiny a3}},{\color{green}{\tiny b1}},{\color{orange}{\tiny b2}}}, 
legend to name=CombinedLegendBar, 
footnotesize,
ybar legend,
group style={group size=3 by 4,vertical sep=65pt
},
height=4.5cm,
width=6.5cm,
ybar=1.5pt,
xtick=data,
tick label style={font=\scriptsize},
symbolic x coords={metric1,metric2,metric3,metric4,metric5,metric6,1,2,3,4,5,6,7,8,9,10,11,12,13,14,15,16,17,18,19,20,21,22,23,24,25,26,27,28},
 ylabel style={align=center}]
 
       \nextgroupplot[title= \text{(a) SWA (c=1, N=1)},ylabel={sustainble throughput}, bar width=2.2pt,xlabel={n (no. of tasks per node)}]
        \addplot[fill=cream,postaction={pattern=north west lines}]  table[x=metric,y=a,col sep=space]  {./PlotData/swg-n.csv};\label{bar1};
        \addplot[fill=aureolin,postaction={pattern=horizontal lines}]  table[x=metric,y=ao,col sep=space]  {./PlotData/swg-n.csv};\label{bar2};
        \addplot[fill=bittersweet,postaction={pattern=north east lines}]  table[x=metric,y=ap,col sep=space]  {./PlotData/swg-n.csv};\label{bar3};
        \addplot[fill=babyblue]  table[x=metric,y=aop,col sep=space]  {./PlotData/swg-n.csv};\label{bar4};
        
        \nextgroupplot[title=\text{(b) YSB (c=1, N=1)},ymin=0,bar width=2.2pt,xlabel={n (no. of tasks per node)}]
         \addplot[fill=cream,postaction={pattern=north west lines}]  table[x=metric,y=a,col sep=comma]  {./PlotData/ysb-n.csv};
        \addplot[fill=aureolin,postaction={pattern=horizontal lines}]  table[x=metric,y=ao,col sep=comma]  {./PlotData/ysb-n.csv};
        \addplot[fill=bittersweet,postaction={pattern=north east lines}] table[x=metric,y=ap,col sep=comma]  {./PlotData/ysb-n.csv};
        \addplot[fill=babyblue]  table[x=metric,y=aop,col sep=comma]  {./PlotData/ysb-n.csv};
        
        \nextgroupplot[title=\text{(c) YSB* (c=1, N=1)},ymin=0,bar width=2.2pt,xlabel={n (no. of tasks per node)}]
         \addplot[fill=cream,postaction={pattern=north west lines}]   table[x=metric,y=a,col sep=comma]{./PlotData/ysbe-n.csv};
         \addplot[fill=aureolin,postaction={pattern=horizontal lines}]  table[x=metric,y=ao,col sep=comma]  {./PlotData/ysbe-n.csv};
          \addplot[fill=bittersweet,postaction={pattern=north east lines}] table[x=metric,y=ap,col sep=comma]  {./PlotData/ysbe-n.csv};
          \addplot[fill=babyblue]   table[x=metric,y=aop,col sep=comma]  {./PlotData/ysbe-n.csv};
        \coordinate (mtop) at (rel axis cs:-2.5,1);


        \nextgroupplot[title=\text{(d) SWA (n=1, N=1)},ylabel={sustainble throughput},bar width=1.79pt,symbolic x coords={1,2,3,3.5,4,5,6,7,8,9,10,11,12,13,14,15,16,17,18,19,20,21,22,23,24,25,26,27,28},xlabel={c (no. of CPUS per task)}]
         \addplot[fill=cream,postaction={pattern=north west lines}]  table[x=metric,y=a,col sep=space]  {./PlotData/swg-c1.csv};
        \addplot[fill=aureolin,postaction={pattern=horizontal lines}]table[x=metric,y=ao,col sep=space]  {./PlotData/swg-c1.csv};
         \addplot[fill=bittersweet,postaction={pattern=north east lines}]  table[x=metric,y=ap,col sep=space]  {./PlotData/swg-c1.csv};
        \addplot[fill=babyblue]  table[x=metric,y=aop,col sep=space]  {./PlotData/swg-c1.csv};
        
        \nextgroupplot[title=\text{(e) YSB (n=1, N=1)},ymin=0,bar width=1.79pt,symbolic x coords={1,2,3,3.5,4,5,6,7,8,9,10,11,12,13,14,15,16,17,18,19,20,21,22,23,24,25,26,27,28},xlabel={c (no. of CPUS per task)}]
         \addplot[fill=cream,postaction={pattern=north west lines}]   table[x=metric,y=a,col sep=comma]  {./PlotData/ysb-c1.csv};
        \addplot[fill=aureolin,postaction={pattern=horizontal lines}]table[x=metric,y=ao,col sep=comma]  {./PlotData/ysb-c1.csv};
        \addplot[fill=bittersweet,postaction={pattern=north east lines}] table[x=metric,y=ap,col sep=comma]  {./PlotData/ysb-c1.csv};
        \addplot[fill=babyblue]  table[x=metric,y=aop,col sep=comma]  {./PlotData/ysb-c1.csv};
        
         \nextgroupplot[title=\text{(f) YSB* (n=1, N=1)},ymin=0,bar width=1.79pt,symbolic x coords={1,2,3,3.5,4,5,6,7,8,9,10,11,12,13,14,15,16,17,18,19,20,21,22,23,24,25,26,27,28},xlabel={c (no. of CPUS per task)}]
         \addplot[fill=cream,postaction={pattern=north west lines}]   table[x=metric,y=a,col sep=comma]  {./PlotData/ysbe-c1.csv};
        \addplot[fill=aureolin,postaction={pattern=horizontal lines}]  table[x=metric,y=ao,col sep=comma]  {./PlotData/ysbe-c1.csv};
         \addplot[fill=bittersweet,postaction={pattern=north east lines}] table[x=metric,y=ap,col sep=comma]  {./PlotData/ysbe-c1.csv};
        \addplot[fill=babyblue]  table[x=metric,y=aop,col sep=comma]  {./PlotData/ysb-c1.csv};
        

    \nextgroupplot[title=\text{(g) SWA (n=4, N=1)},ylabel={sustainble throughput},bar width=2.1pt,xlabel={c (no. of CPUS per task)}]
          \addplot[fill=cream,postaction={pattern=north west lines}]  table[x=metric,y=a,col sep=comma]  {./PlotData/swg-c4.csv};
       \addplot[fill=aureolin,postaction={pattern=horizontal lines}]table[x=metric,y=ao,col sep=comma]  {./PlotData/swg-c4.csv};
       \addplot[fill=bittersweet,postaction={pattern=north east lines}] table[x=metric,y=ap,col sep=comma]  {./PlotData/swg-c4.csv};
        \addplot[fill=babyblue]  table[x=metric,y=aop,col sep=comma]  {./PlotData/swg-c4.csv};
        
          \nextgroupplot[title=\text{(h) YSB (n=4, N=1)},bar width=2.1pt,xlabel={c (no. of CPUS per task)}]
        \addplot[fill=cream,postaction={pattern=north west lines}]  table[x=metric,y=a,col sep=comma]  {./PlotData/ysb-c4.csv};
        \addplot[fill=aureolin,postaction={pattern=horizontal lines}] table[x=metric,y=ao,col sep=comma]  {./PlotData/ysb-c4.csv};
        \addplot[fill=bittersweet,postaction={pattern=north east lines}] table[x=metric,y=ap,col sep=comma]  {./PlotData/ysb-c4.csv};
        \addplot[fill=babyblue]  table[x=metric,y=aop,col sep=comma]  {./PlotData/ysb-c4.csv};
        
        \nextgroupplot[title=\text{(i) YSB* (n=4, N=1)},bar width=2.1pt,xlabel={c (no. of CPUS per task)}]
        \addplot[fill=cream,postaction={pattern=north west lines}]   table[x=metric,y=a,col sep=comma]  {./PlotData/ysbe-c4.csv};
       \addplot[fill=aureolin,postaction={pattern=horizontal lines}]  table[x=metric,y=ao,col sep=comma]  {./PlotData/ysbe-c4.csv};
        \addplot[fill=bittersweet,postaction={pattern=north east lines}] table[x=metric,y=ap,col sep=comma]  {./PlotData/ysbe-c4.csv};
        \addplot[fill=babyblue]  table[x=metric,y=aop,col sep=comma]  {./PlotData/ysb-c4.csv};

  \nextgroupplot[title=\text{(j) SWA (c=7, n=4)},ylabel={sustainble throughput},bar width=2.1pt,xlabel={N (no. of Nodes)}]
  	\addplot[fill=cream,postaction={pattern=north west lines}] table[x=metric,y=a,col sep=comma]  {./PlotData/swgN.csv};
       \addplot[fill=aureolin,postaction={pattern=horizontal lines}] table[x=metric,y=ao,col sep=comma]  {./PlotData/swgN.csv};
       \addplot[fill=bittersweet,postaction={pattern=north east lines}] table[x=metric,y=ap,col sep=comma]  {./PlotData/swgN.csv};
        \addplot[fill=babyblue]  table[x=metric,y=aop,col sep=comma]  {./PlotData/swgN.csv};
        
 \nextgroupplot[title=\text{(k) YSB (c=7, n=4)},bar width=2.1pt,xlabel={N (no. of Nodes)}]
         \addplot[fill=cream,postaction={pattern=north west lines}]   table[x=metric,y=a,col sep=comma]  {./PlotData/ysbN.csv};
       \addplot[fill=aureolin,postaction={pattern=horizontal lines}] table[x=metric,y=ao,col sep=comma]  {./PlotData/ysbN.csv};
        \addplot[fill=bittersweet,postaction={pattern=north east lines}]  table[x=metric,y=ap,col sep=comma]  {./PlotData/ysbN.csv};
        \addplot[fill=babyblue]  table[x=metric,y=aop,col sep=comma]  {./PlotData/ysbN.csv};
        
         \nextgroupplot[title=\text{(l) YSB* (c=7, n=4)},bar width=2.1pt,xlabel={N (no. of Nodes)}]
        \addplot[fill=cream,postaction={pattern=north west lines}]  table[x=metric,y=a,col sep=comma]  {./PlotData/ysbeN.csv};
        \addplot[fill=aureolin,postaction={pattern=horizontal lines}] table[x=metric,y=ao,col sep=comma]  {./PlotData/ysbeN.csv};
       \addplot[fill=bittersweet,postaction={pattern=north east lines}]  table[x=metric,y=ap,col sep=comma]  {./PlotData/ysbeN.csv};
        \addplot[fill=babyblue]  table[x=metric,y=aop,col sep=comma]  {./PlotData/ysbeN.csv};
    \coordinate (mbot) at (rel axis cs:1,0);

\end{groupplot}

\path (mtop|-current bounding box.north)--
      coordinate(legendpos)
      (mbot|-current bounding box.north);
\matrix[
    matrix of nodes,
    anchor=south,
    draw,
    inner sep=0.2em,
    draw
  ]at([yshift=1ex]legendpos)
  {\ref{bar1}& AIR&[5pt]
    \ref{bar2}& AIR-O&[5pt]
    \ref{bar3}& AIR-P&[5pt]
    \ref{bar4}& AIR-OP\\};
%
%
\end{tikzpicture}
\end{figure*}
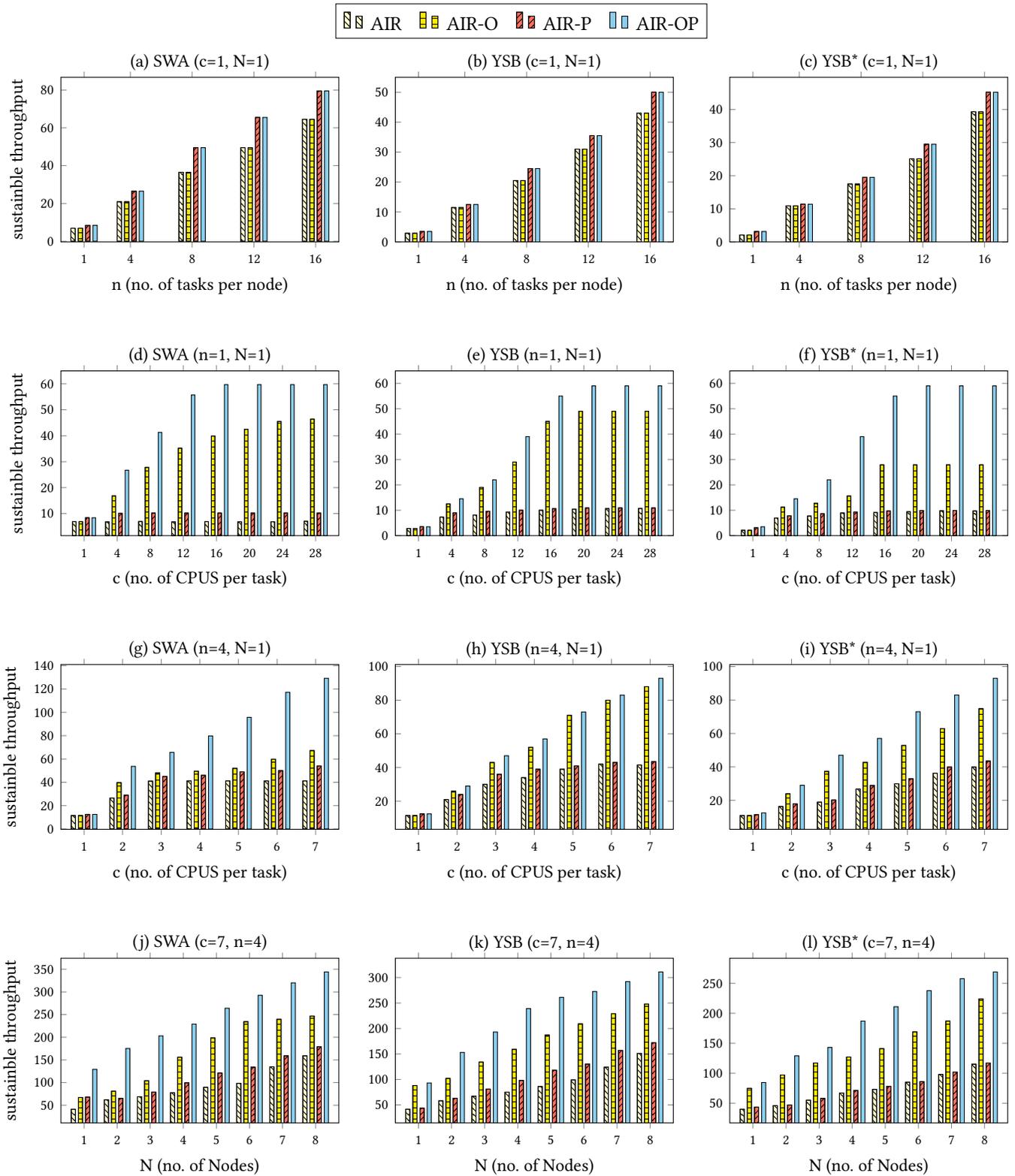

\definecolor{amber}{rgb}{1.0, 0.75, 0.0}
\definecolor{neuesrot}{RGB}{207, 103, 102}
\definecolor{neuesrotup}{RGB}{107, 203, 202}
\definecolor{neuesrot2}{RGB}{117, 113, 102}
\definecolor{neuesrot3}{RGB}{17, 143, 202}
\definecolor{amber2}{RGB}{12, 103, 102}
\definecolor{deepchampagne}{rgb}{0.98, 0.84, 0.65}

\begin{figure*}
\caption{ST values (in $10^6$ events/sec.) of AIR, Spark and Flink over multiple nodes: SWA, YSB and YSB*\label{plot2}}

\begin{minipage}[c]{1\textwidth}
\centering
\begin{tikzpicture}[scale=1][font=\footnotesize]
\begin{axis}[
x tick label style={/pgf/number format/1000 sep=},
ylabel=sustainable throughput,
width=17.5cm, 
height=4.5cm,
ybar=1.5pt,
ylabel near ticks,
xtick=data,
ymin=0,
xlabel=N (no. of Nodes),
symbolic x coords={1,2,3,4,5,6,7,8},
legend style={at={(0.23,.973)},
anchor=north,legend columns=-1},
nodes near coords,
    every node near coord/.append style={font=\tiny},
   nodes near coords align={vertical},
bar width=12.65pt
]

\addplot [color=black, fill=babyblue] coordinates {
(1,129.2) 
(2,175.2) 
(3,203)
(4,229)
(5,264)
(6,292.5)
(7,320)
(8,344)
};

\addplot [color=black, fill=amber,postaction={pattern=north west lines}]  coordinates {
(1,55.498402) 
(2,61.134191) 
(3,73.567344)
(4,74.284703)
(5,74.997501)
(6,76.003250)
(7,75.900600)
(8,76.775637)
};

\addplot [color=black, fill=deepchampagne,postaction={pattern=north east lines}]  coordinates {
(1,48.000000) 
(2,60.500000) 
(3,63.500000)
(4,69.750000)
(5,72.900000)
(6,83.500000)
(7,94.500000)
(8,112.003400)
};
\legend{AIR-OP, SPARK, FLINK  
}
\end{axis}
\node[above,font=\small] at (current bounding box.north) {(i) SWA};
\end{tikzpicture}
\end{minipage}\hfill

\begin{minipage}[c]{1\textwidth}
\centering
\begin{tikzpicture}[scale=1][font=\footnotesize]
\begin{axis}[
x tick label style={/pgf/number format/1000 sep=},
ylabel=sustainable throughput,
width=17.5cm, 
height=4.5cm,
ybar=1.5pt,
ylabel near ticks,
xtick=data,
ymin=0,
xlabel=N (no. of Nodes),
symbolic x coords={1,2,3,4,5,6,7,8},
legend style={at={(0.23,.973)},
anchor=north,legend columns=-1},
nodes near coords,
    every node near coord/.append style={font=\tiny},
   nodes near coords align={vertical},
bar width=12.65pt
]

\addplot [color=black, fill=babyblue] coordinates {
(1,93) 
(2,153) 
(3,193)
(4,239)
(5,261)
(6,272.5)
(7,292)
(8,311)
};

\addplot [color=black, fill=amber,postaction={pattern=north west lines}] coordinates {
(1,5.96) 
(2,12.50) 
(3,27.50)
(4,39.00)
(5,49.52)
(6,52.00)
(7,63.00)
(8,77.62)
};

\addplot [color=black, fill=deepchampagne,postaction={pattern=north east lines}] coordinates {
(1,16) 
(2,21.50) 
(3,38.50)
(4,61.75)
(5,66.00)
(6,75.00)
(7,89.50)
(8,98.00)
};
\end{axis}
\node[above,font=\small] at (current bounding box.north) {(ii) YSB};
\end{tikzpicture}

\end{minipage}\hfill
\begin{minipage}[c]{1\textwidth}
\centering
\begin{tikzpicture}[scale=1][font=\footnotesize]
\begin{axis}[
x tick label style={/pgf/number format/1000 sep=},
ylabel=sustainable throughput,
width=17.5cm, 
height=4.5cm,
ybar=1.5pt,
ylabel near ticks,
xtick=data,
ymin=0,
xlabel=N (no. of Nodes),
symbolic x coords={1,2,3,4,5,6,7,8},
legend style={at={(0.23,.973)},
anchor=north,legend columns=-1},
nodes near coords,
    every node near coord/.append style={font=\tiny},
   nodes near coords align={vertical},
bar width=12.65pt
]

\addplot [color=black, fill=babyblue] coordinates {
(1,84.5) 
(2,129) 
(3,143)
(4,187)
(5,211)
(6,237.8)
(7,257.8)
(8,269)
};

\addplot [color=black, fill=amber,postaction={pattern=north west lines}] coordinates {
(1,6.8) 
(2,10.52) 
(3,22.10)
(4,32.35)
(5,46.00)
(6,51.00)
(7,59.60)
(8,62.30)
};

\addplot [color=black, fill=deepchampagne,postaction={pattern=north east lines}] coordinates {
(1,8) 
(2,13.50) 
(3,28.50)
(4,41.75)
(5,56.00)
(6,61.00)
(7,69.50)
(8,78.00)
};

\end{axis}
\node[above,font=\small] at (current bounding box.north) {(iii) YSB*};
\end{tikzpicture}
\end{minipage}\hfill

\end{figure*}
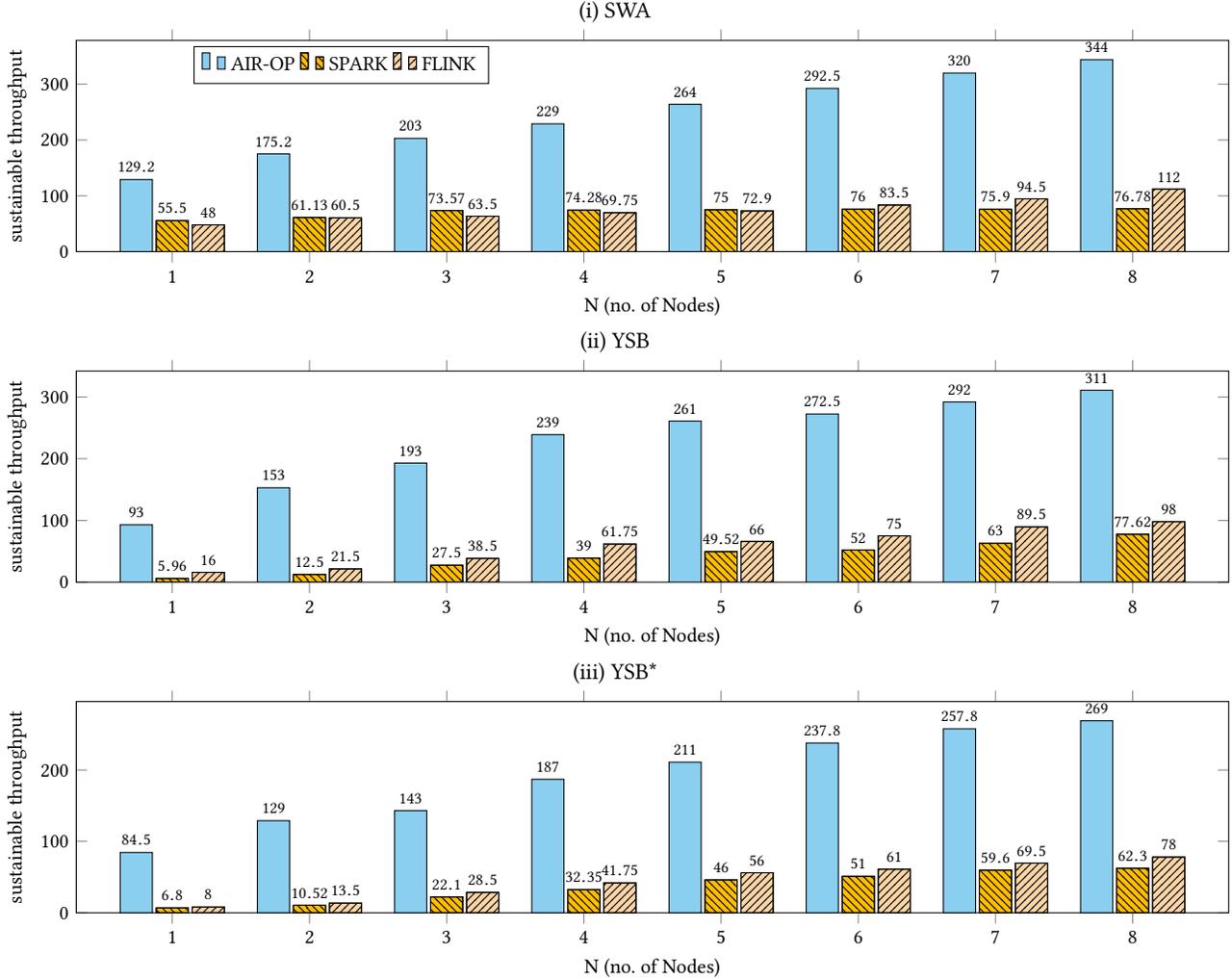
\subsection{Benchmark Setting I: Different AIR Configurations}

\smallskip\noindent\textbf{Setup.~}
This setting exclusively investigates the behaviour of AIR under different configurations and workloads. Depending on whether pipelining (P) and overcommit (O) are enabled or not, we have four variants of AIR, namely AIR, AIR-P, AIR-O and AIR-OP.  These four variants of AIR are evaluated against the three use-cases described in Sections~\ref{sec:swa}--\ref{sec:ysb*} under each choice of SLURM's {\tt -N}, {\tt -n} and {\tt -c} parameters.
First, we study the impact of the {\tt -n} parameter (the number of tasks) by varying it from 1 to 16 and by keeping both {\tt -c} and {\tt -N} fixed to 1. Second, the impact of the number of CPU cores per task, {\tt -c}, is studied by varying it from 1 to 28 and by fixing both {\tt n} and {\tt -N} to 1. Third, to find the best combination of {\tt -n} and {\tt c} for a single node, we performed our experiments with various combinations of these two parameters, thus   
dividing the 28 cores of a compute node among the tasks.  After having determined the best combination of {\tt -n} and {\tt -c} that yields the highest ST value for a single node, we used this combination for allocating similar resources on {\tt -N} compute nodes and thereby determined the ST of the system for multiple compute nodes.

\smallskip\noindent\textbf{Results.~}
As shown in Figures~\ref{fig:metrics} (a)--(c), by varying the setup from 1 task per node to 16 tasks on a single node and with 1 CPU core per task, ST increases by a factor of nearly 9.5 for all the four variants of AIR. The same pattern is observed across all the three use-cases. The plots also show that the overcommit option has no significant impact when the task allocations have only a single core assigned to them. 

Figures~\ref{fig:metrics} (d)--(f) show the impact of the number of CPU cores per task allocation. By keeping the number of tasks per node constantly at 1, we increase the number of CPU cores from 1 to 28. For AIR and AIR-P, it is observed that beyond a particular number of cores, increasing the number of cores has no major impact on the performance. That is, supplying even more cores to a single AIR instance does not increase ST.  It thus becomes evident that the key to further improve is to initialize more AIR instances, each associated with a their optimal amount of CPU cores. For choosing such a setup, we have two options, either using a large number of small (in terms of CPU cores) task allocations at the SLURM level and mapping one AIR instance to each such task allocations, or having fewer task allocations on which we can then allocate multiple AIR instances in overcommit mode. The former option of having a large number of task allocations limits the number of AIR instances to the number of available CPU cores per compute node. In our case, it is typically beneficial to have more than 28 parallel AIR instances per node since not all cores are constantly utilized at 100\%. In addition, for complex use-cases involving lots of asynchronous operators (each running several threads corresponding to the channels they serve), it is generally not beneficial to allocate just one CPU core per task allocation to yield the best performance.

Figures~\ref{fig:metrics} (g)--(i) provide another insight into CPU utilization for 4 tasks and 1 node. We could consider 4 combinations of {\tt -n} and {\tt -c} to effectively divide the CPU cores for performing task allocations: ($n=1$, $c=28$), ($n=2$, $c=14$), ($n=4$, $c=7$) and ($n=7$, $c=4$). The first two combinations could clearly exploit the overcommit option (AIR-O and AIR-OP) to deploy more parallel AIR instances per task allocation. Figures~\ref{fig:metrics} (d)--(f) show that when the overcommit option is enabled, the performance improves nearly 6 times. However, the number of parallel AIR instances on a single task allocation using overcommit cannot be supported beyond a certain number (in our case it was 6) per task allocation. To have more parallel instances deployed on a given node, we thus have to increase the number of task allocations per node while still making sure that each AIR instances is assigned to the best number of CPU cores. In our experiment, the combination ($n=4$, $c=7$), gave us the best ST values for all the three use-cases---refer to plots (g)--(i).

In Figures~\ref{fig:metrics} (j)--(l), finally, we keep $c=7$ and $n=4$ fixed and vary the number of nodes via {\tt -N} from 1 to 8. It is observed that all the 4 modes of AIR are scaling out very well for the three use-cases. As expected, AIR-O and AIR-OP exhibit the best ST values for all the given node counts. 

\smallskip\noindent\textbf{Impact of Pipelining.~} On a single compute node, enabling pipelining (P) while not considering the overcommit (O) option achieves up to 23\% (on avg. 13.3\%) higher ST values under the SWA use-case. On the other hand, using both pipelining and overcommit increases performance by up to 68\% (on avg. 48\%). Across all 8 nodes, we have very similar observations. In SWA, pipelining can be employed only at the {\tt Generator} operator, where the messages can be directly copied to the succeeding {\tt Aggregation} operator. 
In the case of YSB, for a single node, the pipelining option can optimize the message passing at three operators: {\tt Generator}, {\tt Filter} and {\tt Static-Join}, which results in up to 17\% (on avg. 9\%) improvement in ST, whereas enabling also overcommit increases the performance by up to another 55\% (on avg. 38\%).  While running on 8 compute nodes, ST further improves by to up to 69\% (on avg. 61\%) and up to 27\% (on avg. 17\%), respectively. 
In the case of YSB*, enabling pipelining in its 4 stateless operators results in an increase of ST by up to 9\% (on avg. 7.6\%), whereas enabling overcommit in addition increases performance by up to 52\% (on avg. 41\%).  While running on 8 compute nodes, ST increases by up to 65\% (on avg. 61\%) for pipelining and by up to 8\% (on avg. 4\%) for both pipelining and overcommit.

\subsection{Benchmark Setting II: Systems under Comparison}

\smallskip\noindent\textbf{Setup.~}
Finally, we compare the ST of AIR under its best parameter setting on a given number of nodes $N$ with the ST of Spark and Flink under the same numbers of nodes. In the case of Spark, we deploy $N$ task-managers (each with 28 cores except one with 26 cores) and 1 master (2 cores) on a configuration of $N$ cluster nodes. Then we vary the number of executors within each task-manager and its parallelism parameters and report the highest ST we can measure. Similarly, for Flink, we deploy $N$ task-managers (workers) and 1 job-manager (master) on $N$ cluster nodes, and vary the number of slots within each task-manager to determine the highest ST. 

\smallskip\noindent\textbf{Results.~}
The results of this setting are depicted in Figure~\ref{plot2}. Here, AIR outperforms Spark by a factor of up to 4.5 and Flink by a factor of up to 3.6 for the SWA use-case. Similarly, for the more complex YSB setting, AIR performs up to 15 times better than Spark and up to 5.8 times better than Flink. For the even more complicated YSB* use-case, AIR outperforms Spark by up to a factor of 12 and Flink by up to a factor of 9 in terms of ST.  We generally observe that Spark loses performances relatively to Flink and AIR when the dataflows become more complex, i.e., when more sliding-window operators are involved, which seems to be a legacy of the BSP architecture of Spark. We finally highlight that we are able to process 269 million events per second in this setting, which corresponds to a plain data throughput of 36.58 GB per second for an average event size of 136 bytes, when using 8 compute nodes and 224 cores for the YSB* use-case.
\vspace{-1.5pt}



\section{Conclusions}
As highlighted recently~\cite{DBLP:journals/pvldb/ZeuchBRMKLRTM19}, the gap between the real-world throughput of complex engines such as Spark or Flink and the theoretical upper bound given by the main-memory bandwidth of a single (centralized) machine still is about two orders of magnitude. We believe that, with the design of AIR, we found a good compromise for a light-weight, reduced design of a DSPS that exhibits good performance and scales well also to larger cluster deployments. The direct integration into SLURM via the MPI libraries makes AIR an ideal backend for common HPC deployments. We intend to extend the capabilities of AIR in particular with respect to its practical usability and thus develop more user-friendly frontends. As for future research topics, we also plan to investigate new mechanisms to include more explicit forms of workload balancing and fault tolerance directly into the master-less architecture of AIR.

\section{Acknowledgement}
This project is funded by the University of Luxembourg, Luxembourg. We would like to express our gratitude to Mr. Sadi Nasib, M. Sc. student, University of Luxembourg,  for his assistance in setup in the benchmarking environment and to Mr. Jeyhun Karimov, Ph.D. student, Technical University of Berlin, for his valuable suggestions on configuring  the Spark benchmark. We also thank the HPC team of the University of Luxembourg for their timely help and support.
\balance


\bibliographystyle{ACM-Reference-Format}
\bibliography{reference.bib}

\end{document}